\documentclass{aa}
\usepackage{epsfig}

\def\etal{{\rm et al.\thinspace}}
\def\eg{{\rm e.g.\ }}

\def\ie{{\rm i.e.\ }}
\def\cf{{\rm cf.\ }}

\def\spose#1{\hbox to 0pt{#1\hss}}
\def\ltsimm{\mathrel{\spose{\lower 3pt\hbox{$\sim$}}
	\raise 2.0pt\hbox{$<$}}}
\def\gtsimm{\mathrel{\spose{\lower 3pt\hbox{$\sim$}}
	\raise 2.0pt\hbox{$>$}}}
\def\Mdot{\hbox{${\dot M}$} \,}

\def\km{{\rm\thinspace km}}
\def\s{{\rm\thinspace s}}
\def\yr{{\rm\thinspace yr}}
\def\kmps{\hbox{${\rm\km\s^{-1}\,}$}}

\def\Msol{\hbox{${\rm\thinspace M_{\odot}}$}}
\def\Msolpyr{\hbox{${\rm\Msol\yr^{-1}\,}$}}

\begin{document}
   
\title{Self-similar evolution of wind-blown bubbles with mass loading 
by hydrodynamic ablation}

\author{J.M. Pittard, T.W. Hartquist, \and J.E. Dyson}

\institute{Department of Physics \& Astronomy, The University of Leeds, 
        Woodhouse Lane, Leeds, LS2 9JT, United Kingdom\\
              email: jmp@ast.leeds.ac.uk}

\offprints{J. M. Pittard, \email{jmp@ast.leeds.ac.uk}}

\date{Received <date> / Accepted <date>}

\abstract{
We present similarity solutions for adiabatic bubbles that are blown by
winds having time independent mechanical luminosities and that are each
mass-loaded by the hydrodynamic ablation of
distributed clumps. The mass loading is `switched-on' at a specified
radius (with free-expansion of the wind interior to this point) and 
injects mass at a rate per unit volume proportional to $M^{\delta} r^{\lambda}$ where
$\delta = 4/3$ (1) if the flow is subsonic (supersonic) with respect to
the clumps. 
In the limit of negligible mass loading a similarity solution found by 
Dyson (\cite{D1973}) for expansion into a smooth ambient medium is 
recovered. The presence of mass loading heats the flow, which 
leads to a reduction in the Mach number of the supersonic 
freely-expanding flow, and weaker jump conditions across
the inner shock. In solutions with large mass loading, it is possible for 
the wind to connect directly to the contact discontinuity without first
passing through an inner shock, in agreement with previous hydrodynamic
simulations. In such circumstances, the flow may or
may not remain continuously supersonic with respect to the clumps.
For a solution that gives the mass of swept-up ambient gas to be less than
the sum of the masses of the wind and ablated material, 
$\lambda \ltsimm -2$, meaning that the exponent of the density profile 
of the interclump medium must be at most slightly positive, with negative  
values preferred. Maximum possible values for the ratio of ablated mass 
to wind mass occur when mass loading
starts very close to the bubble center and when the flow is supersonic
with respect to the clumps over the entire bubble radius.
Whilst mass loading always reduces the temperature of the shocked wind,
it also tends to reduce the emissivity in the interior of the
bubble relative to its limb, whilst simultaneously increasing the central
temperature relative to the limb temperature. The maximum temperature in
the bubble often occurs near the onset of mass loading, and in some cases 
can be many times greater than the post-inner-shock temperature.
Our solutions are potentially relevant to a wide range of astrophysical 
objects, including stellar wind-blown bubbles, galactic winds, starburst
galaxy superwinds, and the impact of an AGN wind on its surrounding 
environment. This work complements the earlier work of Pittard \etal 
(\cite{PDH2001}) in which it was assumed that clumps were evaporated 
through conductive energy transport.
\keywords{hydrodynamics -- shock waves -- stars: mass-loss -- 
ISM:bubbles -- galaxies: active}
}

\titlerunning{Self-similar evolution of mass-loaded WBBs}
\authorrunning{Pittard et al.}

\maketitle

\label{firstpage}

\section{Introduction}
\label{sec:intro}
Numerous observations of wind-blown bubbles (WBBs) have led to the
conclusion that their structures and evolution are significantly affected
by mass entrainment from embedded clumps. As a case in point, bubbles 
blown by the winds of highly evolved stars can be mass-loaded from
clumps of stellar material ejected during prior stages of massloss 
(Smith \etal \cite{SPDH1984}; Hartquist \etal \cite{HDPS1986}; 
Meaburn \etal \cite{MNBDW1991}; Dyson \& Hartquist \cite{DH1992};
Hartquist \& Dyson \cite{HD1993}). The properties of winds mass-loaded by
material from clumps (or stellar sources) has also received theoretical
attention in the context of ultracompact H II regions (Dyson \etal
\cite{DWR1995}), globular cluster 
winds (Durisen \& Burns \cite{DB1981}), galactic winds and starburst 
galaxy superwinds (Strickland \& Stevens \cite{SS2000}), and accretion flow 
structures at the centers of active galactic nuclei (David \& Durisen 
\cite{DD1989}; Toniazzo \etal \cite{THD2001}).

Investigations into the self-similar nature of supernova remnant (McKee \&
Ostriker \cite{MO1977}; Chi\`{e}ze \& Lazareff \cite{CL1981}; Dyson \&
Hartquist \cite{DH1987}) and stellar wind-blown bubble
(Pittard \etal \cite{PDH2001}) evolution in
tenuous media with embedded clumps have been performed. Detailed 
one-dimensional, time-dependent hydrodynamic models of specific WBBs
associated with evolved stars have also been constructed (Arthur \etal
\cite{ADH1993},~\cite{ADH1994},~\cite{AHD1996}). In at least some isothermal 
mass-loaded winds, Arthur \etal (\cite{ADH1994}) and Williams \etal 
(\cite{WHD1995}, \cite{WDH1999}) showed that a global shock does not
occur in the region where the Mach number is large. Instead, a global shock
occurs only after the wind has travelled far enough for mass loading to 
decelerate it sufficiently that its Mach number is less than a few. In
some of these cases, a global shock within the mass loading region does not
exist at all, and the wind continues until it encounters a termination
shock caused by the interaction of the wind with an external medium.

Observations supporting mass loading in WBB include the detection of
blue-shifted absorption features of species with a range of ionization
potentials in the spectrum of the central star of the Wolf-Rayet nebula
RCW~58. Smith \etal (\cite{SPDH1984}) argued that the observed velocity
spread for the detected features is much larger than would be expected
for $T \ltsimm 10^{5}$~K gas consisting only of stellar wind material
decelerated through a terminal shock, and suggested that the velocity
spread originates due to mass loading of the shocked wind through the
entrainment of material from clumps. Crucially, one can obtain an estimate
for the ratio of wind mass to entrained clump mass: the observed velocity
spread implies a value of 40 or 50 to one. A detailed time-dependent, 
hydrodynamic model including nonequilibrium ionization structure is
consistent with these conclusions (Arthur \etal \cite{AHD1996}).
Spectroscopic data also provide evidence for transonic flows in the halo
of core-halo planetary nebula (Meaburn \etal \cite{MNBDW1991}),
again consistent with high mass loading rates. Finally, it is also 
suspected that
the gradual deceleration of a wind by mass loading and the associated
weakening of an inner shock may in fact contribute to the radio quietness
of some WBBs, although this is a conjecture which currently cannot be proven
(see Williams \etal \cite{WHD1995}, \cite{WDH1999}).
  
In Pittard \etal (\cite{PDH2001}, hereafter PDH), similarity solutions were 
derived for the structures and evolution of mass-loaded WBBs, under the 
assumption that conductive evaporation from embedded clumps was the dominant 
mass loading process, and that both evaporation from the cold swept-up 
shell and radiative losses were negligible. To obtain a similarity solution
with these conditions, specific radial power-laws on the clump and 
interclump density distribution, and temporal power-laws on the wind 
mass-loss rate and terminal velocity, were required. Approximate
similarity solutions for evaporatively mass-loaded WBBs with the assumptions 
of constant mass-loss rate and wind velocity, and an isobaric shocked 
wind region have previously been obtained by Weaver \etal (\cite{WMCSM1977};
evaporation from the cool swept-up shell) and Hanami \& Sakashita 
(\cite{HS1987}; mass-loading from clumps). A central assumption in both
of these papers was that the shocked wind was approximately isobaric. 
However, PDH demonstrated that this was not necessarily a good assumption
(\eg see their Fig.~4). Indeed, imposing this condition is likely to 
set a limit to the amount of mass loading that can occur. 

A central conclusion of PDH was that there exist maximum possible values
for the ratio of evaporated mass to stellar wind mass, as a consequence of
the evaporation rates dependence on temperature and the lowering of
temperature by mass loading. In particular it was difficult to find ratios
approaching what was observationally required.
The work in this paper complements PDH by considering the case in which
hydrodynamic ablation is the dominant mass addition process. As conductively
driven evaporation has a very temperature sensitive rate, ablation is
likely to regulate clump dispersal into lower temperature media.

Our solutions are again potentially relevant to such diverse objects as 
WBBs created by a faster wind interacting with a 
clumpy AGB superwind, by the wind of a young star interacting 
with surrounding molecular material, and the wind of an active galactic 
nucleus impacting its environment.

\section{Similarity Solutions}
\label{sec:sim_solutions}
The basic physics behind the hydrodynamic ablation of material from 
dense clumps into the surrounding flow was presented by Hartquist \etal
(\cite{HDPS1986}). It was proposed that if the flow around a clump
is subsonic, hydrodynamic mixing occurs as a result of the well-known
Bernoulli effect. For flow with a Mach number $M$ with respect to the clump
of less than unity,
this leads to a volume mass injection rate, $\dot{\rho}$, proportional 
to $M^{4/3}$. For supersonic flow, mixing occurs largely as a result
of a low pressure region over the reverse face of the clump, 
`shadowed' from the flow. In this case the mass injection rate is Mach 
number independent.

In this paper we present similarity solutions for WBBs mass-loaded by
hydrodynamic ablation. Our solution method is very similar to that used by 
PDH, and we refer the reader to that paper for a discussion of
many of the assumptions involved. As a starting point, we also assume the 
same qualitative shock structure in the WBB as shown in Fig.~1 of PDH: 
that is a central wind source surrounded by a region of
unshocked wind, separated from an outer region of shocked 
wind by an inner shock. The swept-up ambient medium is 
assumed to radiate efficiently and collapse to a negligibly thin shell 
coincident with the contact discontinuity separating the stellar and ambient 
gas (\cf Dyson \& de Vries \cite{DV1972}).

A central and fundamental difference between the conductive case 
considered in PDH and the ablative case examined here, however, 
is that here mass loading 
may {\em also} occur in the unshocked stellar wind. The effect is to
heat this part of the flow, which leads to 
reduced Mach numbers, and weaker jump conditions across the inner shocks.
We shall see that for high mass loading rates, it is possible for the 
unshocked mass-loaded wind to connect directly with the contact 
discontinuity, without the presence of an inner shock. 

In this work we assume that the flow can be treated as a single fluid.
This requires that the ablated clump material merges with the global flow
in the sense that its temperature, velocity, and density approach those of
the surrounding tenuous material. Ablation by itself might require that the 
clump material mix microscopically with the wind to reach the density and 
temperature of the wind. However, though the mass-loss rate may be controlled
by ablation, thermal conductivity will almost certainly be responsible for
material, once it is stripped from a clump, reaching the physical state of
the tenuous fast flowing medium. Thermal conduction accomplishes this
phase transition without microscopic mixing, and acceleration to the global
flow speed is effected by the response of stripped material to pressure 
gradients and viscous coupling, which may arise from a host of mechanisms
including turbulence. Thus we envisage a two-stage process in which ablation
controls the rate at which mass is stripped from the clump but conductivity
becomes important for the merging into the global flow. 

This receives support
from the comparison of the conductively driven evaporation time of a spherical
clump with $nT = 10^{7}\;{\rm cm^{-3}}$ and $T=10^{4}$~K, embedded in a 
medium with the same pressure (which is typical of planetary nebulae,
Wolf-Rayet wind-blown bubbles, and starburst superwinds) and $T=10^{7}$~K
(which is also typical of hot material in such regions) to the sound
crossing time in the clump. The sound crossing time is somewhat (but not
hugely) smaller than the conductively driven evaporation time for a large 
range of clump sizes, if the Cowie \& McKee (\cite{CM1977}) estimate of the
mass-loss rate driven by saturated conduction is used. Consequently, ablation
initiates mass-loss because it causes an increase in the surface area of a
clump triggering more conductive heat transfer. Use of the analysis of
Cowie \& McKee (\cite{CM1977}) and McKee \& Cowie (\cite{MC1977}) shows 
that for the assumptions given above, radiative losses do not hinder 
conductively driven evaporation unless the clump radius is greater than 
$\approx 15/n_{f}$~pc where $n_{f}$ is the number density (${\rm cm^{-3}}$) 
of the global tenuous flow. Of course, a clump that does not cool
radiatively after it is compressed by a shock will have a shorter sound 
crossing time and be ablated more rapidly. However, by assuming a clump
temperature of $10^{4}$~K above we have established that the two-stage
process is likely to be relevant in many environments if a clump radiatively
cools after being shocked.

Observations of WBBs and PNe indicate that mass loading may not 
necessarily begin at zero distance from the wind source. For
instance, in the young nebula PN Abell 30 the clumps appear almost all 
of the way down to the central star (Borkowski \etal \cite{BHT1995}), 
whilst in  the more evolved Helix Nebula the clumps are all located closer 
to its edge (\eg Meaburn \etal \cite{MCBW1996}). We therefore include
the radius at which mass-loading `switches on' as a free parameter in our 
models, and assume that the wind undergoes free expansion interior to this
radius. Hence each solution will consist of a region of supersonic wind 
with no mass loading and an adjacent region with mass loading.

One can imagine two possible causes for this minimum `mass-loading' radius. 
In one scenario the clumps could have been ejected at low velocity 
from the central star at an earlier evolutionary stage. The ejection of clumps
then abruptly stopped, so that at the time of observation they had
travelled a finite distance from the central star. By this process a 
central region clear of clumps surrounded by a clumpy region can be 
generated. A second possibility is that clumps interior to the mass-loading 
radius have been completely destroyed by the wind. It seems reasonable to 
suppose that clumps located closest to the central star will be 
destroyed first, since they will have been subjected to the wind from the
central star for the longest time. Then as the bubble or nebula evolves, 
clumps at ever increasing distance from the central star will be destroyed.
Timescales for the destruction of clumps by ablation can be estimated from
Hartquist \etal (\cite{HDPS1986}) and Klein \etal (\cite{KMC1994}). Estimated 
destruction timescales vary from significantly less than to greater than
the age of the bubble/PNe, in accord with the different spatial 
distribution of clumps in objects of differing age.

Regardless of which of the above scenarios is responsible for the 
existence of such a minimum mass-loading radius, this radius will 
physically increase with time. Our similarity solution requires that it 
increases in the same way as that of the contact discontinuity 
(\ie $r \propto t^{2/(5+\lambda)}$, where $\lambda$ is the radial 
dependence of the mass-loading). For most of the solutions presented in this
paper, the minimum mass-loading radius scales with or close to $t$. Since on
physical grounds we might expect it to scale as $t$, our solutions
closely match this requirement.

In our solutions an inner shock may or may not be present - in the
latter case the mass-loaded wind directly connects to the contact 
discontinuity, and the mass loading may be strong enough for the wind to 
become subsonic with respect to the clumps before the contact discontinuity
is reached. If an inner shock is present, the postshock flow is by definition
subsonic with respect to the shock, but may still be supersonic with respect
to the clumps. In this case a number of different profiles for the Mach 
number are possible before the contact discontinuity is reached.

At the center of the bubble prior to the onset of mass loading,
we solve only the continuity and momentum equations, with the implicit
assumption that the thermal energy of the flow is negligible, whilst in 
the mass loading regions we additionally solve the energy equation, and 
include a source term for mass injection in the continuity equation. For
a $\gamma = 5/3$ gas, the equations for the mass-loaded flow are:

\begin{equation}
\frac{\partial \rho}{\partial t} + \nabla \cdot (\rho u) = \dot{\rho}
\end{equation}

\begin{equation}
\frac{\partial (\rho u)}{\partial t} + \nabla \cdot (\rho u^{2}) 
 + \frac{\partial P}{\partial r} = 0 
\end{equation}

\begin{equation}
\frac{\partial (\frac{1}{2} \rho u^{2} + \varepsilon)}{\partial t} 
 + \nabla \cdot (\rho u(\frac{1}{2} u^{2} + \frac{5}{3} 
 \frac{\varepsilon}{\rho})) = 0 
\end{equation}

\noindent In these equations the symbols have their usual meanings. In the 
next sections we discuss appropriate similarity variables for these equations,
our treatment of the boundary conditions, and the scaling relationships
to normalize the resulting solutions. The reader is again referred to
PDH for a more in-depth discussion of the details.

\subsection{The Similarity Variables}
\label{sec:ablate}
Let the interclump ambient medium have a density of the form 
$\rho = \rho_{0} r^{\beta}$, and let us consider the case in which the 
mass-ablation rate is also radially dependent: $\dot{\rho} = Q 
 r^{\lambda} M^{4/3}$ for subsonic ablation ($M<1$), and $\dot{\rho} = Q 
 r^{\lambda}$ for supersonic ablation ($M>1$) (\cf Hartquist \etal 
\cite{HDPS1986}). A similarity solution demands that 
$\lambda = (2\beta - 5)/3$, and the `physical' parameters $r, \rho, u$ and
$\varepsilon$ may be expressed in terms of the 
dimensionless similarity variables $x$, $f(x)$, $g(x)$, and $h(x)$ where

\begin{equation}
r = x Q^{-1/(5 + \lambda)} \dot{E}^{1/(5 + \lambda)}  t^{2/(5 + \lambda)}
\end{equation}

\begin{equation}
\rho = Q^{5/(5 + \lambda)} \dot{E}^{\lambda/(5 + \lambda)}
       t^{(5 + 3\lambda)/(5 + \lambda)} f(x)
\end{equation}

\begin{equation}
u    = Q^{-1/(5 + \lambda)} \dot{E}^{1/(5 + \lambda)}
       t^{-(3 + \lambda)/(5 + \lambda)} g(x)
\end{equation}

\begin{equation}
\varepsilon = Q^{3/(5 + \lambda)} \dot{E}^{(2 + \lambda)/(5 + \lambda)}
       t^{-(1 - \lambda)/(5 + \lambda)} h(x).
\end{equation}

\noindent Upon substituting the above similarity variables, the hydrodynamic 
equations for the region of freely expanding wind become:

\begin{eqnarray}
\label{eq:ode1_ab_fe}
& \left(g - \frac{2x}{5 + \lambda}\right)f' + fg' + \frac{2fg}{x} + 
\frac{5 + 3\lambda}{5 + \lambda}f = 0 & \\ 
\label{eq:ode2_ab_fe}
& \left(g - \frac{2x}{5 + \lambda}\right)f' + 
\left(2f - \frac{2xf}{(5 + \lambda)g}\right)g'  + \frac{2fg}{x} + 
\frac{2 + 2\lambda}{5 + \lambda}f = 0 & 
\end{eqnarray}

\noindent where a prime denotes derivation with respect to x. 
For the flow in the mass-loaded region we obtain:

\begin{eqnarray}
\label{eq:ode1_ab}
& \left(g - \frac{2x}{5 + \lambda}\right)f' + fg' & + \frac{2fg}{x} + 
\frac{5 + 3\lambda}{5 + \lambda}f \nonumber \\  
 & & - x^{\lambda} M^{4/3} = 0 \\
\label{eq:ode2_ab}
& \left(g - \frac{2x}{5 + \lambda}\right)g' + \frac{2}{3f}h' & - 
\frac{3 + \lambda}{5 + \lambda}g \nonumber \\ 
 & & + \frac{g}{f} x^{\lambda} M^{4/3} = 0 \\
\label{eq:ode3_ab}
& \left(g - \frac{2x}{5 + \lambda}\right)h' + \frac{5}{3}hg' & - 
\frac{1 - \lambda}{5 + \lambda}h \nonumber \\ 
 & & + \frac{10}{3}\frac{hg}{x} - \frac{g^{2}}{2} x^{\lambda} M^{4/3} = 0 
\end{eqnarray}

\noindent where the Mach number, $M = g \sqrt{9f/(10h)}$. It is 
simple to rearrange these equations to find $f'$, $g'$, and $h'$ which may
then be integrated to obtain solutions.

\subsubsection{Boundary Conditions}
\label{sec:bound_con}
The values of $f$ and $g$ at $x=\Delta x$, the 
inner boundary of our integration, are $g = \sqrt{2/\phi}$ and 
$f = 1/(2 \pi \, \Delta x^{2} g^{3})$, where $\phi$ is a parameter which
specifies the relative amount of mass loading in the solution. 
$h$ is set to zero at this point. Equations~\ref{eq:ode1_ab_fe} 
and~\ref{eq:ode2_ab_fe} are integrated to the radius at which mass loading 
begins. From this point onwards we then integrate 
Equations~\ref{eq:ode1_ab}-\ref{eq:ode3_ab} instead. 
The Mach number of the wind
rapidly decreases from its initial value of infinity and both the density 
and velocity of the wind respond to the mass addition. Integration proceeds  
until the contact discontinuity (CD) is reached. This occurs when the flow 
velocity is equal to the coordinate velocity (\ie $v=dR/dt$),
which is where $g(x=x_{cd}) = 2x/(5+\lambda)$. As previously discussed, this 
can sometimes occur {\em before} the specified position of the inner shock
($x = x_{is}$) is reached, in which case the mass-loaded flow connects
directly to the CD. If an inner shock is present, its
velocity with respect to the bubble center is 
$g_{s} = 2x_{is}/(5+\lambda)$, and the Mach number of the preshock flow 
with respect to the shock is

\begin{equation}
M_{1} = (g_{1}-g_{s}) \sqrt{\frac{f_{1}}{\gamma (\gamma-1) h_{1}}}
\end{equation}

\noindent where the subscript `1' indicates the preshock values. 
Because (unlike for the evaporative case) $M_{1}$ is not necessarily 
large, we cannot assume that
the strong shock jump conditions apply and therefore must use the full
Rankine-Hugoniot relations. The postshock values are then:

\begin{equation}
f_{2} = \frac{\gamma+1}{\gamma-1}\frac{M^{2}}{M^{2}+2}f_{1}
\end{equation}

\begin{equation}
g_{2} = x_{IS} + (g_{1}-x_{IS})\frac{f_{1}}{f_{2}}
\end{equation}

\begin{equation}
h_{2} = \frac{2 \gamma M^{2} - (\gamma-1)}{\gamma+1} h_{1}.
\end{equation}

\noindent Even if the flow velocity has not converged to the
coordinate velocity, an inner shock may not exist. This can occur if the 
mass loading interior to the proposed position of the inner shock is 
strong, resulting in $M_{1}$ being less than unity. In such a scenario, the 
jump conditions are not applied and the integration proceeds as before.

At the CD, we must also satisfy conservation of momentum. We take

\begin{equation}
\label{eq:theta_form}
\theta \equiv \frac{Q^{-1/(5+\lambda)} \dot{E}^{1/(5+\lambda)}}
{\rho_{0}^{-1/(5+\beta)} \dot{E}^{1/(5+\beta)}}
\end{equation} 

\noindent as a measure of the ablated mass to the mass in the swept-up
shell. It follows that at the contact discontinuity,

\begin{equation}
\theta = \left(\frac{(\gamma-1) h}{x^{\beta}\left[g^{2} - \frac{3+\lambda}
{(5+\lambda)(3+\beta)} xg\right]}\right)^{1/(5+\beta)}. 
\end{equation} 

\noindent The rate of massloss from the star is 

\begin{equation}
\label{eq:mdotc}
\dot{M}_{c} = \lim_{x \rightarrow 0} 4 \pi \, x^{2} \, 
f g \, \Xi
\end{equation}

\noindent where 

\begin{equation}
\Xi = Q^{2/(5+\lambda)} \, 
\dot{E}^{(3+\lambda)/(5+\lambda)} \, t^{(6+2\lambda)/(5+\lambda)},
\end{equation}

\noindent which differs from the definition in the conductively driven
case.
The mass loading parameter $\phi$, which is a measure of the ratio of
ablated mass to wind mass is given by $\phi = 
\dot{M}_{c}/\Xi$.

\begin{table*}
\begin{center}
\caption{The influence of $x_{ml}$ on solutions with $\lambda = -3$. Two 
cases are considered: minor mass loading ($\phi$ = 100) whereby the 
value of $x_{ml}$ is relatively unimportant, and appreciable mass loading
($\phi$ = 10) where $x_{ml}$ exerts a major influence.
From left to right, the columns
indicate: i) the value of $x_{ml}$, ii) the ratio of the shock radii and CD, 
iii-vi) the fractions of energy that are thermal and kinetic 
in the bubble, kinetic in the swept-up shell, and radiated, 
vii) the value of $\theta$, 
viii) the ratio of mass within the bubble to the total mass injected by 
the wind (values greater than unity indicate mass loading), ix) the ratio 
of mass in
the swept-up shell to mass in the bubble, x) the preshock Mach number
of the unshocked wind.}
\label{tab:xml_var}
\begin{tabular}{llllllllll}
\hline
$x_{\rm ml}$ & $x_{\rm is}/x_{\rm cd}$ & $IE_{\rm b}$ & $KE_{\rm b}$ & 
$KE_{\rm sh}$ & $E_{\rm rad}$ & $\theta$ & $\Phi_{\rm b}$ & 
$M_{\rm sh}/M_{\rm b}$ & $M_{1}$ \\
\hline
$\lambda=-3, \phi = 100$ & & & & & & & & & \\
0.1  & 0.271 & 0.46 & 0.014 & 0.469 & 0.059 & 102 & 1.454 & 376 & 4.79 \\
0.2  & 0.268 & 0.46 & 0.013 & 0.469 & 0.059 & 101 & 1.371 & 390 & 4.79 \\
0.33 & 0.266 & 0.46 & 0.013 & 0.469 & 0.059 & 99.7  & 1.289 & 408 & 5.00 \\
0.5  & 0.268 & 0.46 & 0.013 & 0.469 & 0.059 & 99.2  & 1.223 & 426 & 5.56 \\
0.9  & 0.264 & 0.46 & 0.013 & 0.469 & 0.059 & 99.0  & 1.152 & 450 & 11.9 \\
\hline
$\lambda=-3, \phi = 10$ & & & & & & & & & \\
0.02 & 0.82 & 0.17 & 0.63 & 0.18 & 0.022 & 2.95 & 6.50 & 0.51 & 1.60 \\
0.1  & 0.76 & 0.22 & 0.53 & 0.22 & 0.028 & 2.95 & 4.48 & 0.80 & 1.45 \\
0.2  & 0.73 & 0.24 & 0.49 & 0.24 & 0.030 & 2.90 & 3.84 & 0.92 & 1.34 \\
0.5  & 0.66 & 0.25 & 0.45 & 0.27 & 0.033 & 2.67 & 2.55 & 1.23 & 1.12 \\
0.9  & 0.67 & 0.24 & 0.39 & 0.33 & 0.041 & 2.85 & 1.75 & 2.17 & 2.53 \\
\hline
\end{tabular}
\end{center}
\end{table*}

\subsection{Scale Transformation and Normalization}
\label{sec:scale_trans}
The dimensionless Eqs.~\ref{eq:ode1_ab}-\ref{eq:ode3_ab} are invariant 
under the following transformation, which we shall call a normalization:

\begin{equation}
\label{eq:norm}
\left.
\begin{array}{ll}
x \rightarrow & \alpha x \\
f \rightarrow & \alpha^{\lambda} f \\
g \rightarrow & \alpha g \\
h \rightarrow & \alpha^{2+\lambda} h 
\end{array}
\right\}.
\end{equation}

\noindent These can be combined to obtain the normalizations for the
mass, and for the kinetic and internal energies of the bubble: 

\begin{equation}
m = 4 \pi \int f x^{2} dx 
 \hspace{15mm} [\alpha^{3+\lambda}] 
\end{equation}

\begin{equation}
k = 4 \pi \int \frac{1}{2} f g^{2} x^{2} dx 
\hspace{10mm} [\alpha^{5+\lambda}] 
\end{equation}

\begin{equation}
i = 4 \pi \int h x^{2} dx 
\hspace{17mm} [\alpha^{5+\lambda}]. 
\end{equation}

\noindent The full equation for the mass in the bubble is: 

\begin{equation}
\label{eq:mbub}
M_{b} = 4 \pi \, \alpha^{3+\lambda} \, \Xi \, t \int^{x_{\rm cd}}_{x=0} 
f x^{2} \, dx = \alpha^{3+\lambda} \frac{\dot{M}_{c} t}{\phi} m.
\end{equation}

\noindent The total mass lost by the star during its lifetime is

\begin{equation}
M_{W} = \int \dot{M}(t) \, t = \frac{5+\lambda}{11+3\lambda} \dot{M}_{c} \, t.
\end{equation}

\noindent The degree of mass loading in the bubble, $\Phi_{b}$, is defined 
as the ratio $M_{b}/M_{W}$. The mass-swept up into the shell,
$M_{sh}$, the ratio of swept-up mass to bubble mass, $M_{sh}/M_{b}$,
the kinetic energy of the shell, $KE_{sh}$, and the $pdV$ work on the
shocked ambient gas, are all identical to the corresponding forms in
PDH. 

\subsection{Solution Procedure}
\label{sec:sol_proc}
If an inner shock exists, three main parameters determine 
the form of a given similarity solution ($\lambda$, $\phi$ and 
$x_{is}/x_{cd}$). However, the ratio of the radius of the onset of 
mass loading to the radius of the inner shock is an additional parameter 
unique to the ablative solutions, and its value ($x_{ml}$) may also
influence the form of the solution. We find that if the mass loading of
the bubble is minor, $x_{ml}$ has relatively little influence, and 
vice-versa (since $\dot{\rho} \propto r^{\lambda}$ and 
$-3 \ltsimm \lambda \ltsimm -2$ for most of the solutions presented,
most of the mass-loading occurs close to the minimum mass-loading radius). 
Alternatively, if an inner shock does not exist, 
$x_{is}/x_{cd}$ and $x_{ml}$ have no meaning as we have defined them. 
In this case we report the ratio of the onset radius of
mass loading to the radius of the contact discontinuity as the parameter
$x_{ml}$. 

The ratio of mass pick-up from the clumps to the mass swept-up from the 
interclump medium (which is related to the value of $\theta$) is
obtained only after a particular solution has been found. For bubbles whose 
evolution is significantly altered by mass loading, we require that 
simultaneously $\phi$ and $\theta$ are both small.

The similarity equations were integrated with a fifth-order accurate
adaptive step-size Bulirsch-Stoer method using polynomial 
extrapolation to infinitesimal step size. Once the CD was reached, 
the similarity variables were rescaled
using the relationships defined in Eq.~\ref{eq:norm} so 
that $x_{cd} =1$. The mass, and kinetic and thermal energies
of the bubble were calculated, as were the kinetic energy of the shell
and the energy radiated from it. The correct normalization to satisfy
global energy conservation was then obtained. Finally, for given values of
$\dot{E}, Q$ and $t$, the similarity variables $x, f, g$ and $h$ may be
scaled into the physical variables $r, \rho, u$ and $\varepsilon$.

\begin{figure*}
\begin{center}
\psfig{figure=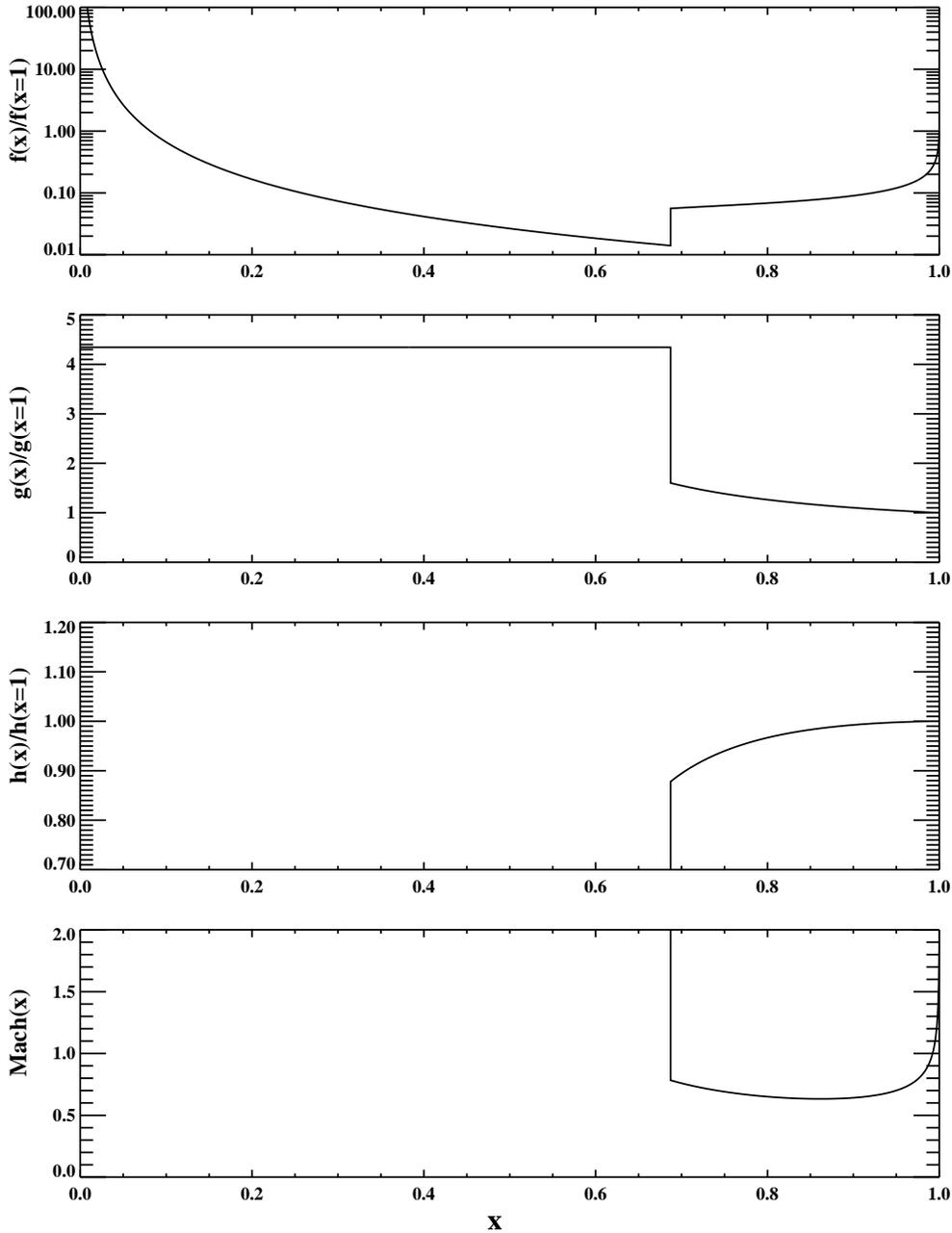,width=13.0cm}
\end{center}
\caption[]{Results for $\lambda = -3$, $x_{\rm is}/x_{\rm cd} = 0.687$, and 
$\phi = 50,000$. In the limit of large $\phi$ (\ie negligible mass loading),
the earlier results of Dyson (\cite{D1973}) are recovered. From the top 
the panels indicate density, velocity, internal energy, and Mach number (with
respect to the stationary clumps) as a function of radius. Except for the
Mach number, the panels are normalized to values of 1.0 at the contact
discontinuity.}
\label{fig:neg_ml}
\end{figure*}
 
\section{Results}
\label{sec:results}

We first checked our work against the solutions obtained by Dyson 
(\cite{D1973}) for WBBs with no mass loading. For this comparison it was 
required that $\lambda = -3$ and that $\phi$ was large. For 
$\lambda = -3$ the radius of minimum mass loading increases as $t$. 
Excellent agreement with Dyson (\cite{D1973}) was found
over a large range of $x_{is}/x_{cd}$. For large $\phi$ (\ie negligible
mass loading) the value of $x_{ml}$ has no effect on the resulting
solution. Fig.~\ref{fig:neg_ml} shows a sample solution.

\begin{figure*}
\begin{center}
\psfig{figure=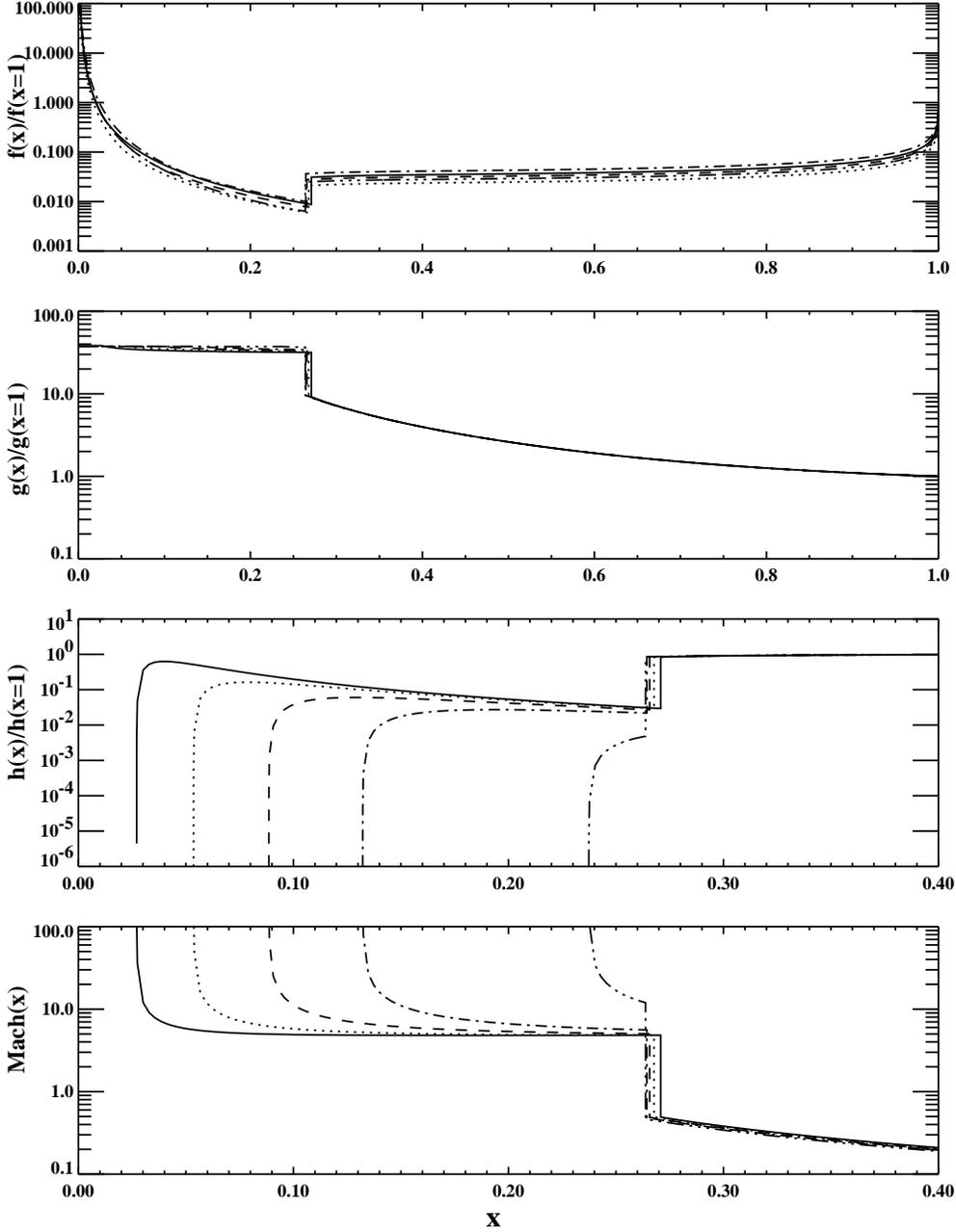,width=13.0cm}
\end{center}
\caption[]{Results for $\lambda = -3$, $\phi = 100$, and various values
of $x_{ml}$. For these parameters the mass loading is small but
non-negligible, and hence demonstrates $x_{ml}$ having a minor influence
on the results. Results for five values of $x_{ml}$ are shown: 0.1 (solid),
0.2 (dots), 0.33 (dashes), 0.5 (dot-dash), and 0.9 (dot-dot-dot-dash). 
The panels are the same as in Fig.~1. In Fig.~3 we 
show solutions where the value of $x_{ml}$ has a much greater influence.}
\label{fig:neg_xml}
\end{figure*}

In Fig.~\ref{fig:neg_xml} we show results for $\lambda=-3$, $\phi=100$,
with different values of $x_{ml}$. The mass loading in these results is 
small, but not negligible, so that the precise value of $x_{ml}$ has 
minor consequences for the
solution obtained. In Table~\ref{tab:xml_var} we tabulate important
parameters from these solutions. In particular, we find that values for 
the ratio $x_{is}/x_{cd}$ and
the energy fractions are very similar. The fractional mass loading of
the bubble, $\Phi_{b}$, is somewhat more sensitive to the value of $x_{ml}$, 
and increases when the `turn-on' radius for mass loading is decreased, as
one would expect. The preshock Mach number also reflects the degree of
preshock mass loading, again as expected. Interestingly, the profiles 
of postshock Mach number are, however, very insensitive to the preshock 
Mach number. With the assumption of constant $\phi$, we note that since
$x_{is}/x_{cd}$ varies for the above results it was important to see 
whether this was a possible factor for some of the differences in these
solutions. Therefore we also investigated the effect of different values 
of $x_{ml}$ whilst keeping $x_{is}/x_{cd}$ constant and varying $\phi$ as 
necessary. We find 
that the resulting solutions are again fairly similar, and differ at about
the same level as the results in the top half of Table~\ref{tab:xml_var}.
Therefore it seems that varying $x_{ml}$ has a direct effect on the
results, and not just an indirect influence through changes induced in
$x_{is}/x_{cd}$ and/or $\phi$.
In the following, therefore, we will vary $x_{ml}$ whilst keeping $\phi$
fixed.

\begin{figure*}
\begin{center}
\psfig{figure=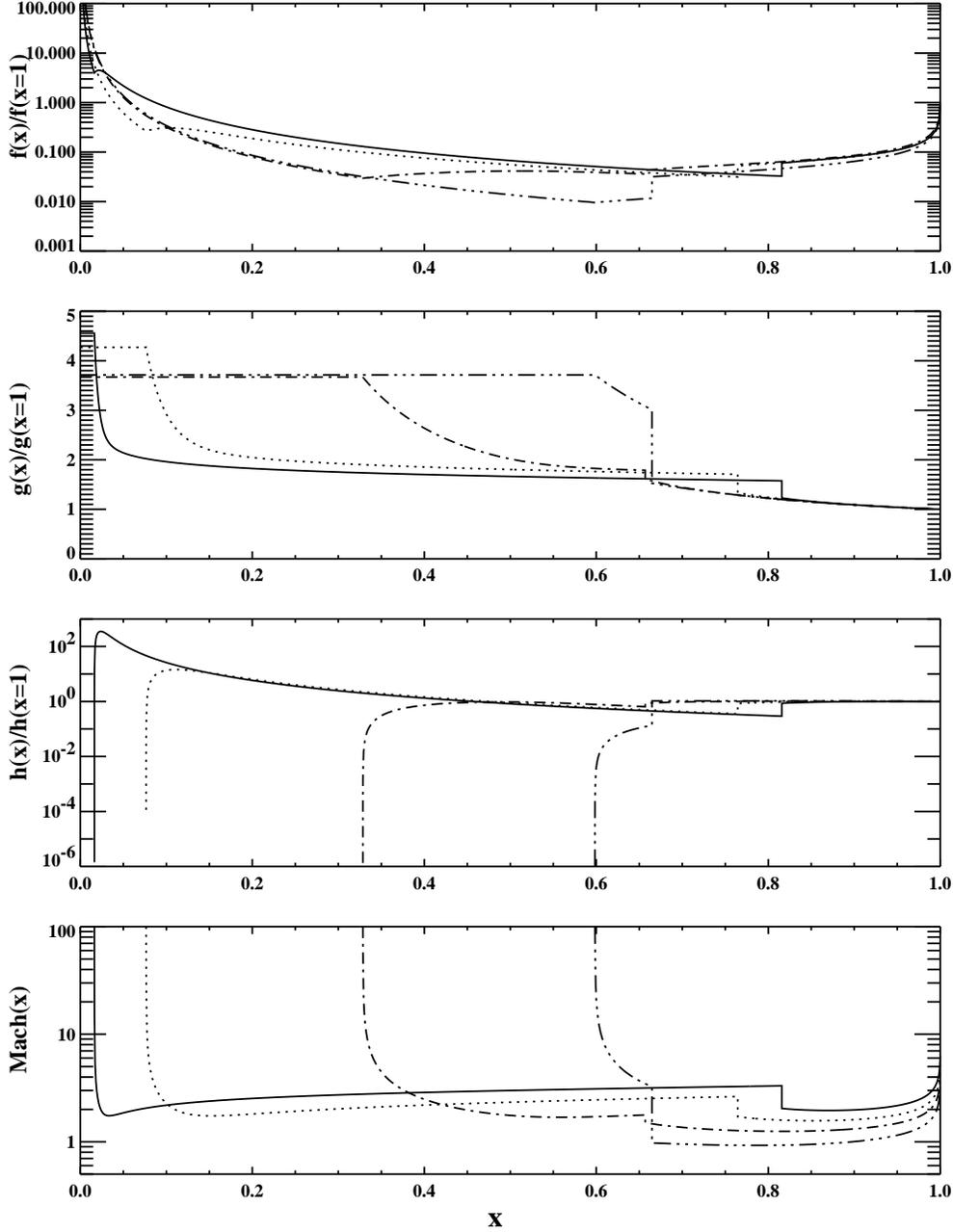,width=13.0cm}
\end{center}
\caption[]{Results for $\lambda = -3$, $\phi = 10$, and various values
of $x_{ml}$. For this value of $\phi$, the mass loading is appreciable,
and hence the solutions shown demonstrate $x_{ml}$ having a major influence.
Results for four values of $x_{ml}$ are shown: 0.02 (solid), 0.1 (dots), 
0.5 (dot-dash), and 0.9 (dot-dot-dot-dash). 
The panels again represent density, velocity, internal energy and Mach number.}
\label{fig:nonneg_xml}
\end{figure*}

At the other extreme, Fig.~\ref{fig:nonneg_xml} shows results where the value
of $x_{ml}$ can have large consequences for the solutions obtained. This 
occurs for values of $\phi$ for which the resulting mass loading of the bubble 
is important. Parameters from these solutions are tabulated in the lower 
half of Table~\ref{tab:xml_var}.
Although the ratio $x_{is}/x_{cd}$ remains relatively insensitive to the
value of $x_{ml}$, the fractional mass loading $\Phi_{b}$, the ratio of the
shell mass to the bubble mass $M_{sh}/M_{b}$, and the energy partition are
all significantly affected, as Table~\ref{tab:xml_var} shows. This is 
expected given that most of the mass loading occurs near the minimum mass 
loading radius for $\lambda = -3$.

\begin{figure*}
\begin{center}
\psfig{figure=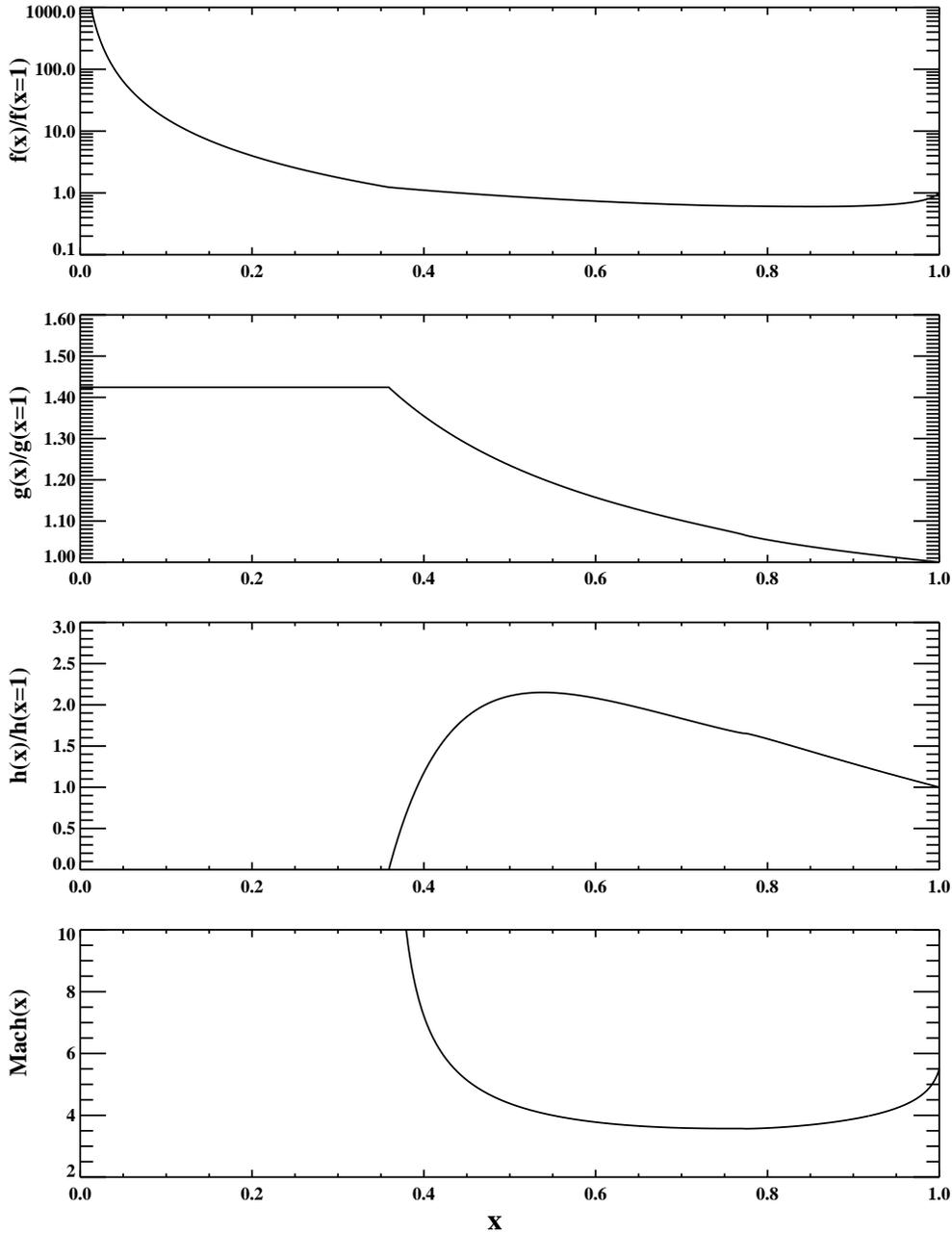,width=13.0cm}
\end{center}
\caption[]{Results for $\lambda = -3$, $\phi = 50$. Here mass loading is
so severe that the flow directly connects to the contact discontinuity. 
For the solution
shown, the onset of mass loading occurs at $x = 0.359 x_{cd}$. Once
the flow starts mass loading, its velocity drops, its relative density 
increases, its kinetic energy drops, and its thermal energy increases.}
\label{fig:direct_con}
\end{figure*}

An interesting consequence of mass loading the wind prior to the innershock
is that if the mass loading is large, the Mach number of the flow can be
reduced so much that the flow directly connects to the contact discontinuity
without the presence of an inner shock. In Fig.~\ref{fig:direct_con} 
we show one solution where this happens. In this example, 
the flow is continuously supersonic with respect to the clumps, and we
obtain $\Phi_{b} = 1.25$ and $M_{sh}/M_{b} = 0.07$. This perhaps indicates
that this morphology is most favoured at early stages of the evolution of 
the bubble before a significant amount of ambient medium is swept-up. In
another example of `direct connection', we find a solution where the Mach 
number of the flow with respect to the clumps drops below unity (see
panel f of Fig.~\ref{fig:machno}). In this case,
$\Phi_{b} = 3.4$ and $M_{sh}/M_{b} = 480$, suggesting a much older bubble. 
We also find that low values of $\lambda$ ($\approx -3$) are apparently 
needed to obtain directly-connected solutions.

\begin{figure*}
\begin{center}
\psfig{figure=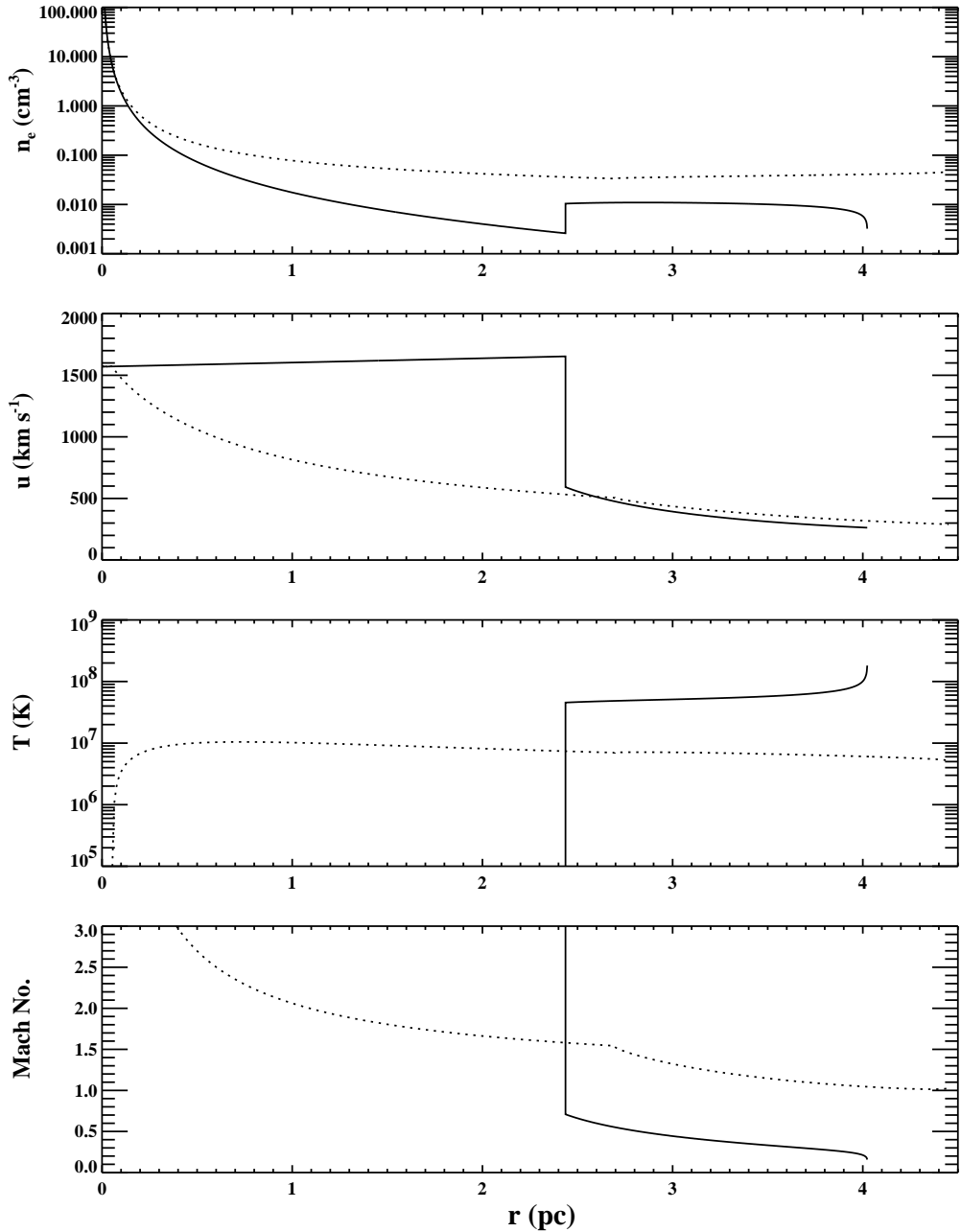,width=13.0cm}
\end{center}
\caption[]{Comparison of solutions with negligible (solid)
and high (dots) mass loading rates, for a given value of $\lambda$ (-2) and
$x_{is}/x_{cd}$ (0.606). The ratio of swept-up mass to bubble mass is
12.3 and 0.96 respectively, whilst the ratio of ablated mass to wind mass is
7.54 for the mass-loaded solution. The preshock Mach number for the 
mass-loaded solution is 1.006. In both cases, the similarity solution has 
been scaled such that the current wind parameters are appropriate for a
WR star: $\Mdot_{c} = 10^{-5} \Msolpyr$ and $v_{w} = 1500 \kmps$.} 
\label{fig:lam-2_MLcomp}
\end{figure*}

In Fig.~\ref{fig:lam-2_MLcomp} we compare solution profiles for the
case where $\lambda = -2$ and for negligible and high mass loading rates.
A $\lambda=-2$ clump distribution is the most physically 
plausible for many astrophysical objects. For this distribution 
the minimum mass loading radius increases as $t^{2/3}$, which is not far 
removed from a linear increase with $t$.  
The solutions have been scaled from the similarity results by assuming 
current values for the wind parameters appropriate for a WR star:
$\Mdot_{c} = 10^{-5} \Msolpyr$ and $v_{w} = 1500 \kmps$. The mass-loss rate
increases with time as $t^{2/3}$ whilst the wind velocity decreases as 
$t^{-1/3}$, so the wind has evolved from a case more suitable to an O-star. 
The assumed bubble age is $t = 10^{4}$~yr. The interclump medium varies
as $\rho \propto r^{-1/2}$ (\ie $\beta = -1/2$) which is a flatter 
distribution than for a constant velocity wind ($r^{-2}$). However, 
since the wind mass-loss rate is increasing and the velocity decreasing, 
an $r^{-1/2}$ interclump density profile is physically realistic for 
such a scenario. 
In comparison with the negligibly mass-loaded solution, the increase in 
density inside the bubble for the highly mass-loaded solution can be 
clearly seen, along with a decrease in preshock velocity, preshock Mach
number, and postshock temperature. Interestingly, the Mach number of the 
postshock flow is {\em higher} for the heavily mass-loaded solution.
Note also that the heavily mass-loaded solution extends to a larger bubble 
radius. This is a consequence of the different preshock interclump densities:
for the heavily mass-loaded solution $n_{e} = 0.066$, whilst $n_{e} = 0.195$
for the negligibly mass-loaded solution.

\begin{figure*}
\begin{center}
\psfig{figure=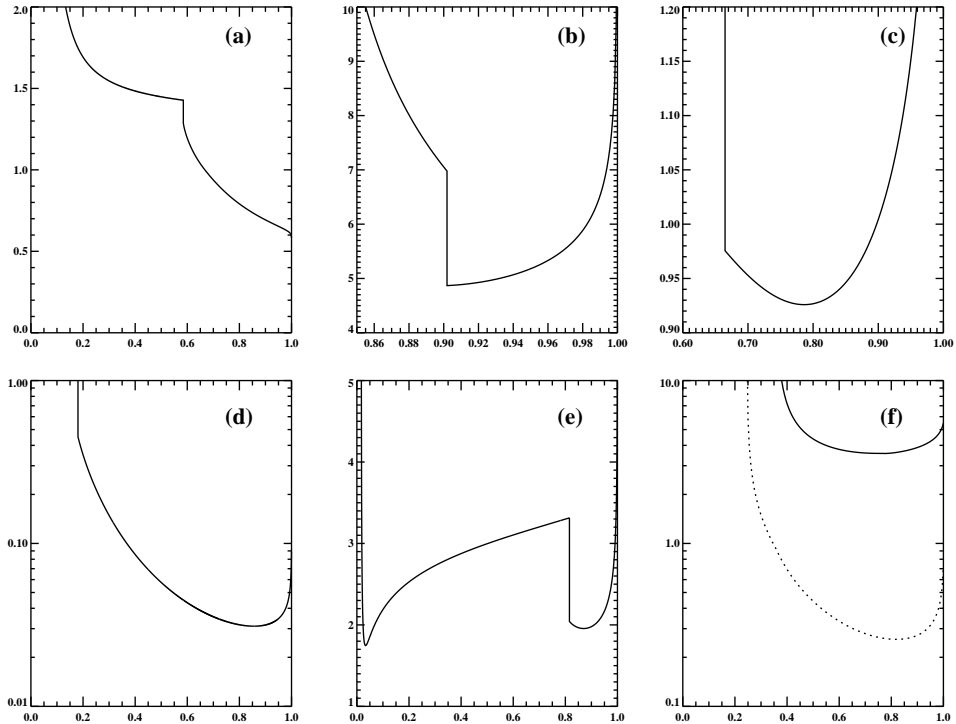,width=13.0cm}
\end{center}
\caption[]{The Mach number as a function of $x$ for a number of different 
solutions. {\em With respect to the clumps}, the shocked region can be either
entirely subsonic, entirely supersonic, or have one (or maybe more) 
sonic points.
The panels show: {\bf a)} shocked region entirely subsonic with a monotonic
decrease in Mach number ($\lambda = -2$, $\phi = 0.5$, $x_{is}/x_{cd} = 
0.585$, $x_{ml} = 0.1$); {\bf b)} shocked region entirely supersonic with 
a monotonic increase in Mach number ($\lambda = -3$, $\phi = 100$, 
$x_{is}/x_{cd} = 0.902$, $x_{ml} = 0.9$); {\bf c)} postshock Mach 
number initially subsonic and decreasing, before levelling off and 
subsequently increasing through a sonic point ($\lambda = 4$, $\phi = 10$, 
$x_{is}/x_{cd} = 0.297$, $x_{ml} = 0.1$); {\bf d)} shocked region entirely 
subsonic with a decreasing Mach number followed by a modest increase close
to the contact discontinuity ($\lambda = -3$, $\phi = 10^{5}$, 
$x_{is}/x_{cd} = 0.181$, $x_{ml} = 0.1$); {\bf e)} postshock flow
continuously supersonic though with an initial decrease in Mach number 
($\lambda = -3$, $\phi = 10$, $x_{is}/x_{cd} = 0.816$,
$x_{ml} = 0.02$); {\bf f)} continuously supersonic flow (solid) directly
connected to the contact discontinuity (\ie no inner shock) 
($\lambda = -3$, $\phi = 50$, $x_{ml} = 0.359$) and a
flow becoming subsonic (dots) directly connected to the contact discontinuity 
($\lambda = -3.2$, $\phi = 12.5$, $x_{ml} = 0.246$).}
\label{fig:machno}
\end{figure*}

In PDH it was shown that there is a 
negative feed-back mechanism caused by the evaporation of mass from 
embedded clumps, which set a maximum limit to the amount that a bubble 
could be massloaded. Additionally, for the bubble mass to be greater than 
the mass of the swept-up shell, a large radial dependence of the 
mass loading was required ($\lambda \geq 4$). In this work we instead
find that small values of $\lambda$ are required to satisfy this condition. 
Our simulations show that it is impossible to obtain a similarity solution 
with $M_{sh} < M_{b}$ for $\lambda \geq -1$ (\eg for $\lambda = 1$, 
$M_{sh} \gtsimm 2.4 M_{b}$). However, for smaller values of $\lambda$, 
solutions satisfying $M_{sh} < M_{b}$ can be found (\eg for 
$\lambda = -2$ (-3), we can obtain 
$M_{b} \approx 8$ (29) $M_{sh}$, where $\Phi_{b} \approx 7.6$ (15)).
For a given value of $M_{sh}/M_{b}$ there appears to be a maximum
value for $\Phi_{b}$. It occurs when mass loading
starts very close to the bubble centre (\ie $x_{ml} \rightarrow 0$) and
when the flow remains supersonic relative to the clumps over the entire
bubble radius.

\begin{figure*}
\begin{center}
\psfig{figure=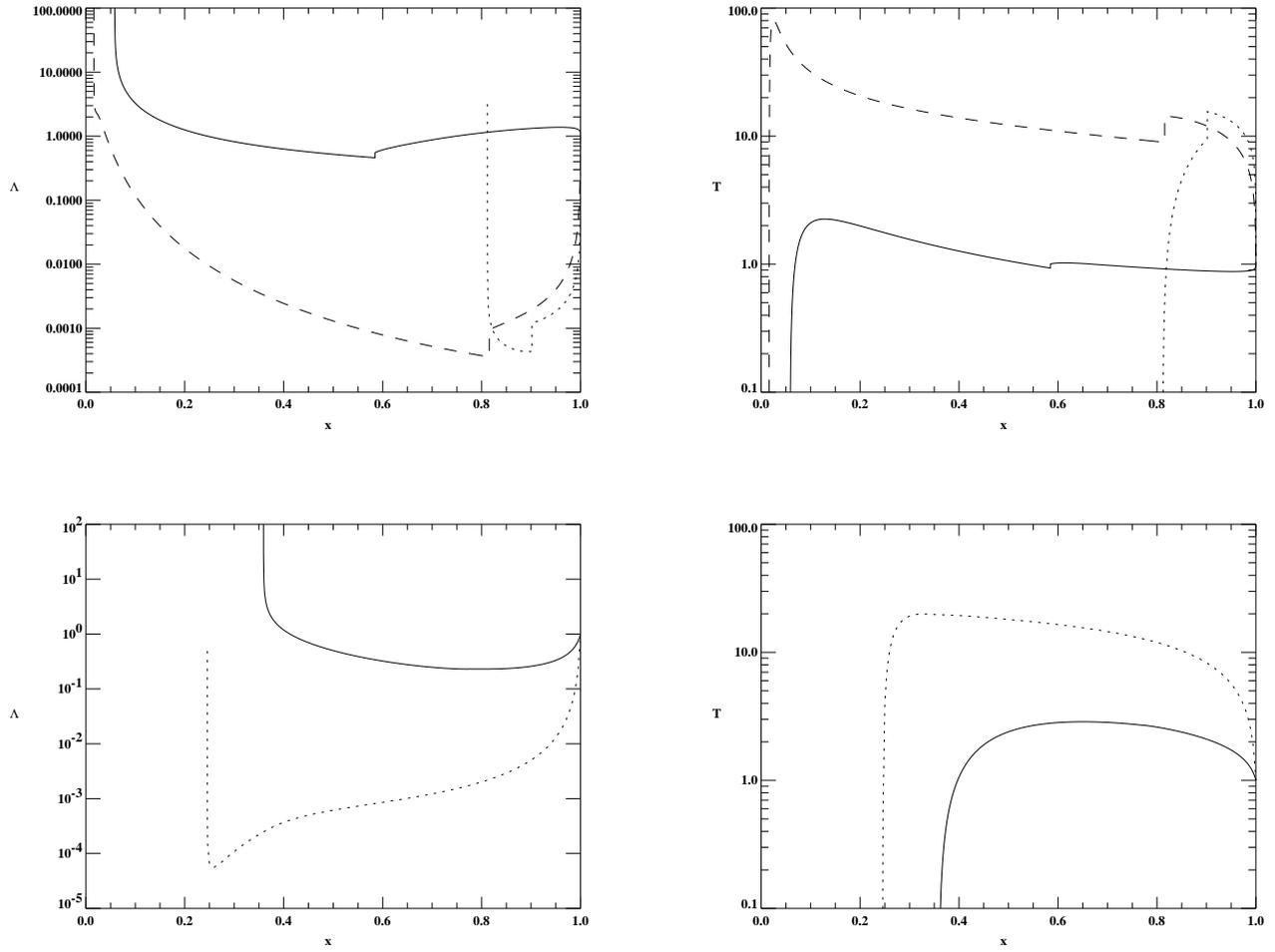,width=18.0cm}
\end{center}
\caption[]{Radial profiles of emissivity per unit volume and temperature,
normalized to values of 1.0 at the limb. The upper panels show results 
where an inner shock is present, whilst the bottom
panels show results where the mass loading region directly connects to the
contact discontinuity. For the upper panels the results shown are for: 
$\lambda = -2$, $\phi = 0.5$,
$x_{is}/x_{cd} = 0.59$, $x_{ml} = 0.1$ (solid); $\lambda = -3$, $\phi = 100$,
$x_{is}/x_{cd} = 0.90$, $x_{ml} = 0.9$ (dots); $\lambda = -3$, $\phi = 10$,
$x_{is}/x_{cd} = 0.82$, $x_{ml} = 0.02$ (dashes). For the lower panels the
results are for: $\lambda = -3$, $\phi = 50$, $x_{ml} = 0.36$ (solid); 
$\lambda = -3.2$, $\phi = 12.5$, $x_{ml} = 0.25$ (dots).}
\label{fig:xray_ab}
\end{figure*}

We have also investigated the Mach number profile of the solutions that
we obtain, of which some are already shown in Figs.~\ref{fig:neg_xml} 
and~\ref{fig:nonneg_xml}. As for the evaporative mass loading in 
PDH, we note that a number of different forms
are possible, and summarize these in Fig.~\ref{fig:machno}.

Finally, radial profiles of the X-ray emissivity per unit volume and 
temperature have been calculated (Fig.~\ref{fig:xray_ab}).
We assume that the emissivity $\Lambda \propto n^{2} T^{-1/2}$, which is a
good approximation over the temperature range $5 \; 10^{5} {\rm K} 
\ltsimm T \ltsimm 5 \; 10^{7}{\rm K}$ (\cf Kahn \cite{K1976}). The upper
panels show results where an inner shock is present, and we will discuss these
first. The solution with the solid line displays relatively constant values
of $\Lambda$ and $T$, and is characterized by a high degree of mass loading
($\Phi_{b} = 9.7$). Note that the highest temperature in the bubble does
not occur postshock, but rather shortly after mass loading begins. The very
high emissivity at the onset of mass loading is a consequence of the low
temperature at this point and the assumed $T^{-1/2}$ dependence of our 
emissivity, and is not physical. The solution with the dotted line has an 
interior dominated by free-expansion of the wind source. Mass loading 
only occurs over the final 20 per cent of the radius, and is 
relatively weak, and the maximum temperature in the bubble occurs 
immediately postshock. The solution with the dashed line has some 
characteristics of each of the other solutions. It is mass-loaded
from small radius which again leads to the maximum bubble temperature
occurring near the center. However, the mass loading is less severe 
($\Phi_{b} = 6.5$) and its radial dependence steeper ($\lambda = -3$) than
for the solution with the solid line, so that there is a large fall in
density between the switch-on radius for mass loading and $x_{is}$. This leads
to a steep rise in emissivity from the inner shock to the bubble center in
a similar fashion to the solution with the dotted line.

The bottom panels of Fig.~\ref{fig:xray_ab} show results where there is no
inner shock and where the mass loading region directly connects to the 
contact discontinuity. The Mach number profiles for these solutions are
displayed in the bottom-right panel of Fig.~\ref{fig:machno}. As 
can be seen, the temperature profiles are approximately flat over their
respective mass loading regions, and are very similar in shape to each other.
This contrasts with the behaviour of their emissivity profiles, where it 
is clear that the solution represented by the solid (dotted) line has 
$\Lambda$ generally decreasing (increasing) with radius.

A general result from Fig.~\ref{fig:xray_ab}, which was also found in the
evaporatively mass-loaded bubbles of PDH, is that higher mass loading 
tends to decrease the central emissivity and increase the
central temperature relative to the limb.

\subsection{Comparison with observations and other theoretical work}
\label{sec:comparisons}
Although our work cannot be directly compared to the hydrodynamic 
simulations of Arthur \etal (\cite{ADH1993},~\cite{AHD1996}), a number 
of similarities exist. These include a sharp increase in temperature 
at the onset of mass loading, and often $\partial \rho/\partial x > 0$ 
over the mass-loaded region. A noticeable achievement of the work of
Arthur \etal (\cite{AHD1996}) was the reproduction of the observed correlation 
between the velocity and ionization potential of ultraviolet absorption 
features towards the central star (Smith \etal \cite{SPDH1984}). Our
solutions also often reproduce this correlation (\eg see 
Fig.~\ref{fig:lam-2_MLcomp} which shows the velocity and temperature
decreasing together for the heavily mass-loaded solution), as do those
of Hanami \& Sakashita (\cite{HS1987}). Interestingly, solutions with
mass-loading from conductive evaporation of the swept-up shell do not 
reproduce this correlation (see Weaver \etal \cite{WMCSM1977}).

\section{Summary}
\label{sec:summary}
We have investigated the evolution of a mass-loaded wind blown bubble
with a constant rate of injection of mechanical energy from a central wind
source. The mass loading occurs due to the 
hydrodynamic ablation of distributed clumps, and is `switched-on' at a 
specified radius, interior to which the wind expands freely.
The requirement that the solution be self-similar imposes a link 
between the radial variation of the interclump density  
($\rho \propto r^{\beta}$) and the rate of mass loading from the clumps 
($\dot{\rho} \propto r^{\lambda}$) which forces $\lambda = (2\beta - 5)/3$.

We first produced solutions with negligible mass loading and $\beta=-2$
(which correspond to constant $\dot{M}_{wind}$ and $v_{wind}$),
which we compared with results obtained by Dyson (\cite{D1973}). 
Excellent agreement for the structure of the bubble was found. 
We also confirmed that for negligible mass loading, the value of $x_{ml}$
had no effect on the resulting solutions, as desired.

We then investigated the changes in the structure of the bubble for 
different values of $\lambda$, $\phi$, $x_{\rm is}/x_{\rm cd}$, and 
$x_{\rm ml}$. The central conclusions are:
\begin{itemize}
\item Substantial mass loading of the wind-blown bubble can occur over
a wide range of $\lambda$. However, to additionally satisfy
the requirement that the bubble mass is larger than the mass of
the swept-up shell, low values of $\lambda$ are needed (\eg $\lambda 
\ltsimm -2$).
\item The profiles of the flow variables are significantly altered under
conditions of large mass loading. With respect to solutions with negligible 
mass loading, the density and temperature profiles increase, and the 
velocity and Mach number profiles drop. Changes to the profiles can be
very rapid. Mass loading of the wind also reduces the Mach number prior
to the inner shock.
\item The mass-loaded wind may also connect directly to the contact 
discontinuity without the need for a global inner shock.
\item The Mach number of the flow relative to the 
clumps which are injecting the mass can take several different forms.
The flow can be either entirely supersonic, or have one (or maybe more) 
sonic points.
\item Whilst mass loading reduces the bubble temperature, it also
tends to reduce the emissivity in the interior of the
bubble relative to its limb, whilst simultaneously increasing the central
temperature relative to the limb temperature. The maximum temperature in
the bubble is often not the post-inner-shock temperature, but occurs near 
the onset of mass loading. In some cases this can be many times greater
than the post-inner-shock temperature.
\end{itemize}

\begin{acknowledgements}
JMP would like to thank PPARC for the funding of a PDRA position, and 
Sam Falle and Rob Coker for helpful discussions. We would also like
to thank an anonymous referee whose suggestions improved this paper. This
work has made use of Nasa's Astrophysics Data System Abstract Service.
\end{acknowledgements}

\end{document}